\documentclass[sigconf]{acmart}

\AtBeginDocument{%
  \providecommand\BibTeX{{%
    \normalfont B\kern-0.5em{\scshape i\kern-0.25em b}\kern-0.8em\TeX}}}
    
\usepackage{framed}

\setcopyright{none}
\renewcommand\footnotetextcopyrightpermission[1]{}

\copyrightyear{2023} 
\acmYear{2023} 
\acmConference[]{To appear in ACM e-Energy}{2023}{ }
\acmBooktitle{ }

\usepackage{xcolor}

\begin{document}

\title[Blockchain-enabled Parametric Solar Energy Insurance via Remote Sensing]{Blockchain-enabled Parametric Solar Energy Insurance\\ via Remote Sensing}

 \author{Mingyu Hao}
 \affiliation{%
   \institution{Australian National University}
   \country{Australia}}
 \email{mingyu.hao@anu.edu.au}

 \author{Keyang Qian}
 \affiliation{%
   \institution{Australian National University}
   \country{Australia}}
 \email{keyang.qian@anu.edu.au}

 \author{Sid Chi-Kin Chau}
 \affiliation{%
   \institution{Australian National University}
   \country{Australia}}
 \email{sid.chau@acm.org}

\begin{abstract}
Despite its popularity, the nature of solar energy is highly uncertain and weather dependent, affecting the business viability and investment of solar energy generation, especially for household users. To stabilize the income from solar energy generation, there have been limited traditional options, such as using energy storage to pool excessive solar energy in off-peak periods or financial derivatives from future markets to hedge energy prices. In this paper, we explore a novel idea of {\em parametric solar energy insurance}, by which solar panel owners can insure their solar energy generation based on a verifiable geographically specific index ({\em surface solar irradiation}). Parametric solar energy insurance offers opportunities of financial subsidies for insufficient solar energy generation and amortizes the fluctuations of renewable energy generation geographically. Furthermore, we propose to leverage blockchain and remote sensing (satellite imagery) to provide a publicly verifiable platform for solar energy insurance, which not only automates the underwriting and claims of a solar energy insurance policy, but also improves its accountability and transparency. We utilize the state-of-the-art succinct zero-knowledge proofs ({\em zk-SNARK}) to realize privacy-preserving blockchain-based solar energy insurance on real-world permissionless blockchain platform Ethereum.

\end{abstract}


\keywords{Solar Energy Insurance, Solar Irradiation Modelling, Blockchain, Remote Sensing, Privacy, Zero-Knowledge Proofs}

\sloppy
\maketitle

\section{Introduction}

Solar energy generation, particularly from household rooftop PV panels, is becoming a major source of renewable energy for a sustainable energy future. But there are many factors affecting the real-life business viability of solar panel deployment. Essentially, the decision of installing a solar panel is a financial problem for many household users, who need to properly factor in the cost and future income of solar energy generation. However, solar energy generation has a lot of uncertainty, including its dependence on weather and climate events like La Ni\~na  phenomenon, affecting the effective surface solar irradiation and solar energy generation.

Traditionally, there are tools for estimating and predicting the solar energy generation at certain geographical locations (e.g., ANN based tool \cite{rodriguez2018predicting}). However, prediction is insufficient to address the financial stability of solar energy generation. 
 Particulatly, for community solar (with up to hundreds of users), even though the impact of solar energy fluctuations is small for one household user, the aggregate impact on a community can be quite large. To stabilize the income from solar energy generation, there have been limited traditional options. For example, one can utilize energy storage to pool excessive solar energy in off-peak periods to maximize solar energy feed-in to the grid. One may also hedge energy prices by financial derivatives from future energy markets on global energy prices. However, these solutions do not directly mitigate the uncertainty of local solar energy generation.

Since solar energy generation is intrinsically correlated with natural events, we draw our inspiration from natural-event-contingent insurance products. For example, crop insurance \cite{kshetri2021blockchain,jha2021blockchain} allows farmers to insure their crop yield against natural events like drought. There are other similar insurance products, like flooding and bushfire insurance \cite{lin2020application}. Therefore, we explore a novel idea of solar energy insurance, by which solar panel owners can insure their solar energy generation based on a certain verifiable geographically specific index (e.g., surface solar irradiation). There are several key advantages of solar energy insurance:

\begin{enumerate}

\item
It offers opportunities of financial subsidies to insufficient solar energy generation, particularly, for disadvantaged communities that rely on local solar energy generation. Solar energy insurance can be bundled as a part of governmental incentive schemes to promote solar panel deployment. Since solar energy insurance is benchmarked with the expected local solar energy generation, it is more effective to address specific users' needs than usual broad-brush incentive schemes like tax rebates.

\item
It can effectively amortize the fluctuations of renewable energy sources geographically, and with other renewable energy sources. Note that different renewable energy sources are affected differently by weather and natural events. We envision a general renewable energy insurance for other renewable energy sources to allow insurance providers to hedge the generation of one source against another. 

\item
It also provides a viable investment opportunity to institutional investors, who seek to diversify their risk by natural-event-based investment. Solar energy insurance, like other natural-event-contingent insurance products, can be securitized into bonds, and be sold in money markets to institutional investors, like pension funds.

\end{enumerate}

In this paper, we introduce the idea of solar energy insurance as non-traditional {\em parametric} insurance. Unlike traditional indemnity insurance that reimburses an insuree according to specific loss in an event based on a case-by-case review, parametric insurance claims can be made based on certain publicly observable parameters (e.g., data measurements from trusted third-party weather services or sensors and satellite imagery). The verification of parametric insurance can be conducted automatically, and insurance payouts can be approved algorithmically without manual intervention. Parametric insurance is commonly provided to crop, flooding and bushfire insurance \cite{jha2021blockchain,lin2020application}, and is also ideal for solar energy insurance.

As a salient feature of parametric insurance, the underwriting and claims of insurance policies can be automated through blockchain and remote sensing (satellite imagery). Blockchain has been recently applied to diverse energy applications (e.g., \cite{CZ22energyplan, ZC21energyplan, WCZ22energyshare, WCZ21energyshare}).
Blockchain helps insurance providers save processing time, cut costs, and comply with regulations. Furthermore, {\em permissionless} blockchain platforms (e.g., Ethereum) ensure transparency/traceability/accountability by an open ledger. In blockchain-based parametric insurance, the insurance provider first codifies the insurance policy by smart contracts. The insuree then submits insurance claims with supporting data (e.g., trusted third-party satellite imagery) to the smart contracts. There are already several parametric insurance providers on permissionless blockchain (e.g., Etherisc \cite{Etherisc}). In this paper, we develop a blockchain-based solution for parametric solar energy insurance.

Blockchain-enabled insurance has the following advantages over non-blockchain indemnity insurance: 
\begin{itemize}

\item
{\bf Verifiability}: Transparency and immutability of distributed ledgers ensure anyone in the system can supervise and verify the correctness of insurance policy execution.

\item
{\bf Automation}: It allows automatic insurance claims without human intervention, resulting in lower administration cost and faster claim speed.

\item
{\bf Easy Assessment}: As parametric insurance does not require case-by-case investigation and adjustment, it minimizes the long-existing problems of traditional insurances such as moral hazard and reverse selection. 
\end{itemize}

But there are two challenges with a blockchain-based solution:
\begin{itemize}

\item {\bf Privacy:} Blockchain does not protect privacy by default, because its data is publicly disclosed on the ledger. Although the insurer knows the insuree's personal information, any outsider who can access smart contracts on blockchain should not know the personal information. Hence, we provide a privacy-preserving solution to conceal private information on smart contracts.

\item {\bf Blockchain Processing Cost:} Permissionless blockchain (e.g., Ethereum) has limited capacity, and its smart contract processing is metered by gas cost. Processing a full insurance policy on blockchain may incur a high gas cost. Hence, we execute only the essential verification of insurance claims on blockchain.

\end{itemize}

\begin{figure}[t]
  \centering
  \includegraphics[width=0.9\linewidth]{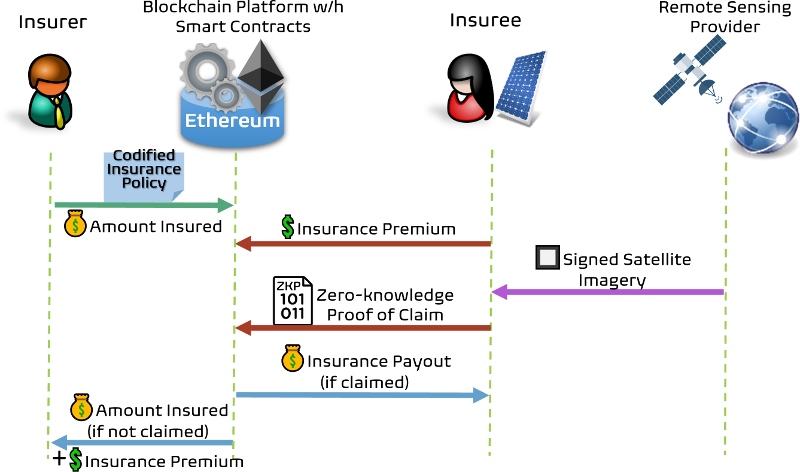}
  \caption{An illustration of blockchain-based solar energy insurance.}
  \label{fig:scenario}
\end{figure}

In our blockchain-based solution, we utilize the state-of-the-art succinct zero-knowledge proofs to realize privacy-preserving blockchain-based solar energy insurance on real-world permissionless blockchain platform Ethereum. Zero-knowledge proofs (zk-SNARK) can  prove the validity of insurance claims which will be verified efficiently by smart contracts.

In Fig.~\ref{fig:scenario}, we describe the idea of blockchain-based solar energy insurance. Before claiming, the insurer sets up the amount insured on blockchain (e.g., Ethereum) by cryptocurrency and the insurance policy by a smart contract. The insuree confirms the policy by paying the insurance premium to the smart contract. To claim the insurance, the insuree first obtains the signed satellite imagery from a trusted remote sensing provider, and then constructs a zero-knowledge proof of claim to prove the correctness of the sources of satellite imagery and claim determination algorithm (based on the computed surface solar irradiation from the satellite imagery). The smart contract verifies the claim on blockchain. If being verified successfully, the insuree will receive the insurance payout.

\section{Backgrounds and Preliminaries}
This section describes the background knowledge of our solution. 

\subsection{Remote Sensing for Solar Irradiation}
The core of parametric insurance is the index determining the payout condition and amount. For solar energy insurance, this corresponds to choosing the measurement of to what extent natural events have affected solar PV generation. Among all other factors, solar radiation, or Surface Solar Irradiation (SSI), has the most significant effect and high variability and forecast uncertainty  \cite{tarigan2014assessment, marquez2013proposed}. Consequently, a key risk of solar PV generation comes from the volatility of SSI, which is used as the basic index of the insurance.

We utilise satellite remote sensing to estimate SSI accurately, effectively, verifiably, and at low cost. Remote sensing can characterize natural features and physical objects on the ground and monitor their changes over time. Compared to on-ground sensors, remote sensing can cheaply collect information over larger spatial areas at various spatial resolutions. Research has shown that SSI estimation based on satellite sensing has higher accuracy and reliability \cite{vindel2016temporal, zhang2016evaluation}. And it is easy to integrate the data with other information to aid decision-making. Moreover, satellite data is usually open-sourced with public accessibility for non-commercial use (e.g., from NASA \cite{USGS} or ESA \cite{EarthOnline} satellites), which enables users to verify the authenticity of the data. Hence a wide range of satellite-based SSI estimation models have been proposed and applied in industry \cite{karlsson2017clara, cebecauer2016site, rigollier2004method, zhang2018estimation}. One of the most well-known statistical models is the Heliosat method \cite{cano1986method}, which is based on the simple fact that the top-of-atmosphere reflectance observed by satellite is statistically proportional to the cloud transmission. Hence, it is possible to derive SSI from cloud observations by satellite imagery.

\vspace{-5pt}
\subsection{Blockchain-based Parametric Insurance}

Blockchain uses distributed ledger technology (DLT) to record transactions. It is decentralized, meaning all participants have a copy of the ledger and can verify on-chain records anytime. The blockchain is immutable, and once a transaction is recorded, as time goes by it is increasingly convinced that the transaction is settled and not retractable. Such properties make the ledger trustful without any centralized manager. A smart contract is an executable program running on blockchain-based on pre-programmed rules (i.e., scripts). Compared to traditional paper-based contracts, smart contracts guarantee the policies specified in code scripts supporting immutability, and no trusted authority is required to monitor the execution of the contract. Because of the transparency of blockchain, anyone in the network can validate the details of the contract. Such properties are ideal for realizing parametric insurance. 

\vspace{-5pt}
\subsection{Succinct Zero-knowledge Proofs}

The goal of Zero-Knowledge Succinct Non-interactive Argument of Knowledge (zk-SNARK)  \cite{snarkintro, Flashproofs} is to enable a prover, who processes a secret, to convince a verifier about the possession of the secret, without revealing the secret itself. The proof is succinct as the verifier can verify the statement in constant time regardless of the input size. And the proof is non-interactive as the prover only needs to communicate with the verifier once. Then the verifier can verify the proof without further communication with the prover. These properties make zk-SNARK applicable for permissionless blockchain platforms like Ethereum. 

One useful initialisation of zk-SNARK is Sonic  \cite{maller2019sonic}, which supports a universal and continually updatable structured reference
string (SRS) that scales linearly in size. Sonic zk-SNARK has a constant size and the marginal cost of verification is comparable with the most efficient zk-SNARKs in the literature.

In Sonic protocol, the prover takes three private input vectors $\mathbf a, \mathbf b, \mathbf c \in \mathbb Z^n_p$ and proves that such vectors satisfy a number of multiplicative and linear constraints. Regardless of the complexity of the constraint system, zk-SNARK protocols like Sonic can verify the correctness of computation with efficient time complexity. Therefore, encoding parametric insurance into the constraint system not only enables zero-knowledge proof of insurance claims but also improves the system's scalability. A more formal description of the constraint system is presented in Appendix \ref{appendix:snark_background}.

\section{MODELS AND FORMULATION}
In this section, we formally define the solar energy parametric insurance model without considering privacy. Privacy and decentralization properties are incorporated into the model in the later sections. We assume all numbers are integers.

\subsection{Remote Solar Irradiation Estimation Model}
We construct solar energy insurance based on remote sensing of solar irradiation. The model is based on the Heliostat-2 method  \cite{rigollier2004method} that converts observations made by geostationary meteorological satellites into estimation of the ground-level solar irradiation. Intuitively, more clouds in the sky results in less sunrise on the ground below. Heliosat-2 first considers a clear-sky model, which estimates the maximum expected irradiation when the sky is clear ($G_{\rm cs}$). Then a ``cloud index'' ($n_t$) is computed based on satellite imagery to capture the amount of cloudiness. Finally, the estimated solar irradiation is a percentage of the clear-sky estimate where the ratio, or clear-sky index ($K_t$), is a segmentation function of the cloud index. 

We describe the deduction of the formula and detailed assumptions in appendix \ref{SSI_detail}. In brief, we assume that the clear-sky models and other coefficients are pre-calculated before the deployment of the model. At the time $t$, the model's input consists of the satellite observed radiance $L_t$ and calibration coefficient $f_t$. Then in the area of one pixel, given $T$ sample points from the target period, the estimated total SSI received during the period is:
\begin{equation}
   G_{\rm gnd}=G_{\rm prd}\sum_{t=1}^{T}(K_{t} G_{{\rm cs}_t})/\sum_{t=1}^{T}G_{{\rm cs}_t}
\end{equation}
where $(G_{{\rm cs}_1}, G_{{\rm cs}_2}, \cdots, G_{{\rm cs}_T})$ are the clear-sky irradiation at each time and $G_{\rm prd}$ is the clear-sky irradiance over the period of the remote sensing data. More details can be found in the Appendix.

\subsection{Decentralised Parametric Insurance Model}
\label{Insur_detail}
In this section, we propose a Parametric Solar Energy Insurance (PSEI) based on the remote sensing model introduced above. The insurance aims to cover the volatility risk of deviation from the baseline solar irradiation value.

Before signing a PSEI contract, the insurer and the insured should agree upon the following parameters:

\begin{enumerate}
\item \textbf{Insured Area \& Insured Duration}: The geometric location and range on which the insurance will cover, as well as the time coverage of the insurance. The area is represented by pixels from satellite images. As estimating actual SSI over the period requires sampling from the period, the insurer also needs to set the sampling frequency and sample size $T$, as mentioned in the previous section. 

\item \textbf{Remote Sensing Provider (RSP)}: The RSP is the source of raw satellite data. We assume that the remote sensing data provider can be a trusted provider with verifiable data (e.g., digital signature from authority) or an oracle provider that aggregates multiple data sources (e.g., Chainlink). We assume that the insurer and insuree can access satellite imagery data with negligible cost. 

\item \textbf{Expected Solar Energy \& Trigger Threshold}: The baseline PV energy $G_e$ is the expected amount of solar irradiation over the insured time and location. The trigger threshold $\epsilon$ is the percentage level at which payment will be triggered if actual solar irradiation falls below.

\begin{figure*}[t]
  \centering
  \includegraphics[width=\textwidth]{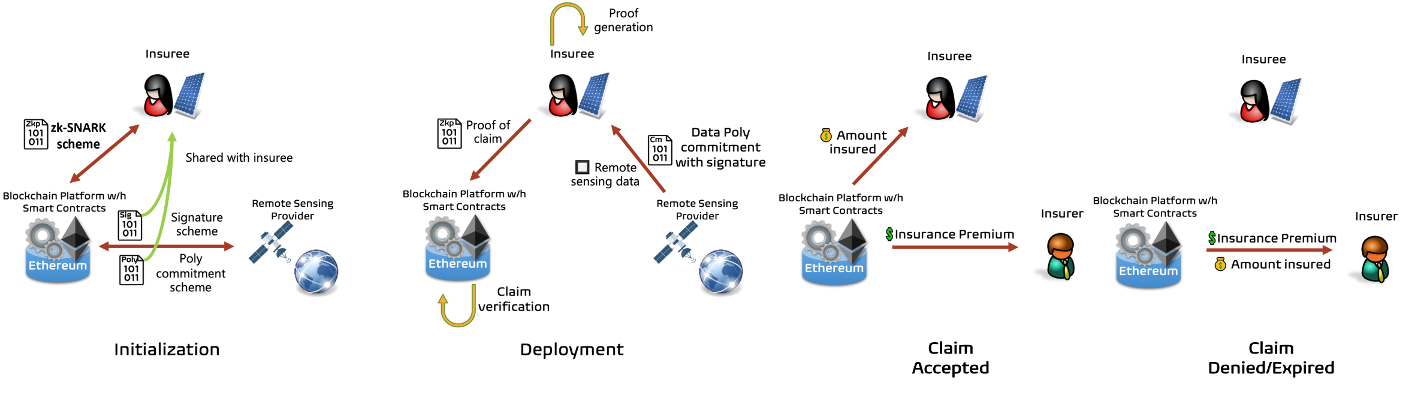}
  \caption{A workflow of the private-preserving solar energy insurance protocol.}
  \label{fig:protocol_flow}
\end{figure*}

\item \textbf{SSI Model Parameters}: Pre-defined SSI model parameters based on historical statistics or geometrical information.

\item \textbf{Other Common Parameters}: The sum insured, premium, payment settlement approaches, expiration date, and other contract terms. 
\end{enumerate}

Once the insurer and insuree agree upon the above parameters, the insurance will be deployed by a smart contract on-chain. However, the contract will not progress until the insured has paid the premium and the insurer has paid the sum insured to the contract. 

Afterwards, The insuree can make an insurance claim at any time. When the insuree wants to claim, the RSP will provide the required model input of the insured area at chosen timestamps as vectors of pixels. Then, the estimated SSI per pixel will be calculated based on the model in the previous section. The estimated total SSI is the sum of SSI over an area. The contract will check whether the estimated solar irradiation is lower than the threshold, namely:
\begin{equation}\label{eq:2}
    G=\sum_{i,j} G_{\rm gnd}(i,j)<G_e\cdot\epsilon
\end{equation}
And if the estimated solar irradiation is indeed less than the threshold, the contract will transfer the sum insured to the insuree's account address. The contract is terminated when the claim is settled or the expiring date is passed. The premium is transferred to the insurer's address before the contract’s termination.

Thanks to the immutability and transparency of blockchain, the terms of the contracts can be automatically verified and executed, and anyone can validate the execution results.

\section{\mbox{Privacy-preserving Insurance Claim}}
In this section, we introduce the privacy-preserving components of the PSEI protocol. We aim to satisfy the following privacy and security requirements:
\begin{enumerate}
    \item \textbf{Private Data Concealment}: The insuree's identity and insurance details (e.g., insured location and time) should not be revealed to any other users.

    \item \textbf{Zero-knowledge Claim}: Claiming the insurance requires satellite images as input for index computation. However, such images also reveal the insured location and time and other private information of the insuree. Therefore, it is necessary to ensure that the claim computation is performed correctly without revealing private data. 

    \item \textbf{False Claim Prevention}: The insuree can be malicious who tries to claim the insurance even though the claim conditions are not satisfied. While preserving honest users' privacy, We also aim to ensure the correctness of the claim process.
\end{enumerate}

We propose a Sonic-style zk-SNARK claim protocol. The key difference from the original Sonic is that both RSP and the client participate in the protocol. The claim protocol can be divided into three stages: Initialization, Deployment, and Payment Settlement. 
\begin{enumerate}

\item
\textbf{Initialization}: Before deployment of the smart contract, public parameters used for the claim process are chosen and shared between insuree, contract, and RSP. Principally this includes two parts: initialisation of the polynomial commitment scheme and signature scheme between RSP and insuree, and initialisation of the zk-SNARK scheme between the smart contract and the insuree. 
As shown in Fig. \ref{fig:protocol_flow}, upon selecting RSP, the RSP sets up a signature scheme and sends the signature public key to the contract. Our protocol adapts to common digital signature schemes such as ECDSA \cite{johnson2001elliptic}. The RSP then setup a polynomial commitment scheme and stores the public parameters on-chain. The contract set up the zk-SNARK scheme and other essential public parameters. Using pre-defined SSI model parameters, the SSI computation model is converted into the Sonic constraints system and verifiable by the zk-SNARK protocol. The detailed construction process is presented in Appendix \ref{appendix:snark_detail}. 

\item
\textbf{Deployment}: After the insuree transfers the premium and the insurer transfers the sum insured to the contract, the smart contract is deployed and waiting for claiming. The policyholder can claim the payment by proving that the claiming condition is satisfied. The claim process has three steps. The concrete protocol is presented in Appendix \ref{Appendix:protocol_details}.

\begin{enumerate}
    \item \textbf{Data Request}: The insuree first requests satellite data from RSP. The RSP responses with raw data, a polynomial commitment of the data, and a signature on the hash of the commitment. 

    \item \textbf{Proof Generation}: The insuree can now generate proof of claim following the zk-SNARK protocol proposed in Appendix \ref{appendix:snark_detail}. The proof mainly includes polynomial commitments and the opening of polynomials at random challenge points specified by the verifier. Because the insuree knows remote sensing data and RSP's polynomial commitment scheme, they can compute openings of the RSP's commitment locally, and the signature signed on RSP's commitment ensures the insurer cannot cheat by providing fake data. 

    \item \textbf{Claim Verification}: Finally, the verifier checks the claim's validity. The verifier first validates the data source by verifying the RSP's signature on polynomial commitment. Then it verifies the claim condition in a zero-knowledge manner following the zk-SNARK protocol and decides whether to accept the claim or not. 
\end{enumerate}

\item
\textbf{Insurance Payout}: Based on the results of claim verification, the insurance payout is settled to the addresses stored in the initialization stage. If the claim is valid, the sum insured would be automatically transferred from the contract account to the insuree's account. The contract is terminated if the claim happens or the expiring date is passed. If the claim fails, the premium will be transferred to the insurer's account before the contract terminates. 

\end{enumerate}

\section{Conclusion}

In this paper, we proposed a novel blockchain-enabled parametric solar energy insurance via remote sensing model, design the required zk-SNARK protocol used in insurance claiming and discuss the extending significance of our model for other renewable energy type insurances and applications.

We presented a zk-SNARK protocol that aims to minimize the incurred gas cost on blockchain and preserve private information of insurance claims. But the implementation and empirical evaluation of our blockchain-enabled solution on Ethereum will be a subject of future study. In future work, we will conduct on-chain experiments to test our protocol. We will also further optimize the on-chain verification computation and off-chain proving computation. We will incorporate blockchain-based solar energy insurance with other community-based energy sharing systems \cite{p2p19,CE20sharing,CE17sharing}. A more general framework of blockchain-enabled parametric insurance protocol can be found in our technical report \cite{techreport}.

\begin{acks}
We would like to thank our shepherd and anonymous reviewers for helpful suggestions. This research was supported by ARC Discovery Project No: GA69027/DP200101985.
\end{acks}

\balance
\bibliographystyle{ACM-Reference-Format}
\bibliography{bibs}

\newpage

\appendix
\section{Remote solar irradiation estimation Model} \label{SSI_detail}
To estimate the solar irradiation from remote sensing, we first estimate the clear-sky solar irradiation based on a clear-sky model from \cite{rigollier2004method} with corrections for the site elevation:
\begin{equation}
   G_{\rm cs} (w_1, w_2) = B_{\rm cs} (w_1, w_2) + D_{\rm cs} (w_1, w_2) 
\end{equation}   
where $w_1$ is the sunrise hour angle,  $w_2$ is the sunset hour angle, $B_{\rm cs}$ is beam irradiation, and $D_{\rm cs}$ is diffusion irradiation. Noticing that such a model is independent of remote sensing, and can be pre-computed before any application. In this paper, we assume the clear-sky irradiation is known and constant. 

Given time $t$ and pixel indexed by $(i, j)$, the cloud index $n_t(i,j)$ is computed by:
    \begin{equation}
    n_t(i,j)=\frac{\rho_t(i,j)-\rho^t_{\rm gnd}(i,j)}{\rho^t_{\rm cloud}(i,j)-\rho^t_{\rm gnd}(i,j)} \\
    \end{equation}
where 
\begin{itemize}

\item 
$\rho_t(i,j)$ is the apparent albedo observed by the spaceborne sensor:
   \begin{equation}
   \rho_t(i,j)=\frac{\pi L_t(i,j)}{I_{\rm 0met}\epsilon(t){\rm cos}(\theta_s(i,j,t))}
   \end{equation}

\item    
$\theta_s$ is the sun zenithal angle, $L$ is the radiance observed by satellites, and $I_{\rm 0met}\epsilon(t)$ is the total irradiation in the visible channel for the various Meteosat sensors with correction used to allow for the variation of sun-earth distance. Because of the fixed positions on the ground, we assume $I_{\rm 0met}\epsilon(t)$ and ${\rm cos}(\theta_s(i,j,t))$ are constants and directly obtained from the satellite, and then we concatenate them into a calibration coefficient $f$. Hence, we simplify the formula of $\rho_t(i,j)$ as:
\begin{equation}
\rho_t(i,j) = f_t(i,j)\cdot L_t(i,j) \\
\end{equation}

\item 
$\rho^t_{\rm cloud}$ and $\rho^t_{\rm gnd}$ are the apparent cloud albedo and the apparent ground albedo respectively. Apparent ground albedo is the apparent reflectance under clear sky estimated as the minimum historical ground albedo under the assumption that there exist instants for which the sky was clear. As mentioned in the Heliosat-2 software description  \cite{lefevre2004description}, the map of ground albedo should be constructed in advance. Therefore, similar to the clear sky model we assume a pre-computed constant apparent ground albedo. 
The apparent cloud albedo is the apparent reflectance of the brightest cloud:
   \begin{equation}
   \rho^t_{\rm cloud}(i,j)=\frac{\rho^t_{\rm eff}(i,j)-\rho^t_{\rm atm}(\theta_s,\theta_v,\psi)}{T(\theta_s)T(\theta_v)} \\
   \end{equation}
where $\rho^t_{\rm eff}$ and $\rho^t_{\rm atm}$ are the effective cloud albedo and intrinsic reflectance of the atmosphere. $T(\theta_s)$ and $T(\theta_v)$ are the global transmittances of the atmosphere for the respective incident and upward radiation. These values depend only upon the satellite zenithal angle and sun zenithal angle. For simplicity, we also assume pre-computed constant apparent cloud albedo. 
\end{itemize}

With these assumptions we now rewrite the cloud index of pixel $(i, j)$ at time $t$ as:
\begin{equation}
n_t(i,j)=\rho_t(i,j) \cdot \sigma_0(i,j) -\sigma_1(i,j) 
\end{equation}
where $\sigma_0$ and $\sigma_1$ are constants. 
With the cloud index $n_t$, we can compute the clear-sky index based on their correlation function. For simplicity, we assume a linear relationship:
   \begin{equation}
   K_{t}(i,j)=1-n_t(i,j)\\
   \end{equation}

Finally, to calculate SSI over a period (e.g., a day, a month), $T$ sample points are chosen and their weighted average clear-sky index is used as the period clear-sky index. Given clear-sky irradiation at each sample points $(G_{{\rm cs}_1}, G_{{\rm cs}_2}, \cdots, G_{{\rm cs}_T})$, and $G_{\rm prd}$ the clear-sky irradiation over the period of the remote sensing data, after computing the respective clear-sky index $(K_{1}, K_{2},\cdots,K_{T})$, the final SSI $G_{\rm gnd}$ is given by:
   \begin{equation}
   G_{\rm gnd}(i,j) =G_{\rm prd}\cdot S_{G_T}^{-1} \cdot \sum_{t=1}^{T}(K_{t}(i,j) \cdot G_{{\rm cs}_t}) 
   \end{equation}
where $S_{G_T}^{-1}=1/\sum_{t=1}^{T}G_{{\rm cs}_t}$. Finally, the total SSI over an area covered by a set of pixels is:
    \begin{equation}
    G = \sum_{i,j} G_{\rm gnd}(i,j)
    \end{equation}

We remark that in this paper we assume proportional correlation between SSI and solar energy generation. However, in reality there could be other factors affecting the resultant energy generation, such as PV panel tilt angle, panel generation efficiency, dust on panel, loss in transmission and storage, etc. Therefore, in the Initialization stage of the insurance contract, the insurer needs to estimate and determine appropriate parameters for the index algorithm by themselves. For example, the insurer can collect sample data from local electricity meters and then compare with estimated values to construct a model for residual error distribution. 

Since SSI is computed using remote sensing images captured by orbit satellites, which usually has revisit time around 1-2 days, the computed SSI could be different from actual SSI due to the discrete nature of sample points. Note that the proposed method does not restrict the number of remote sensing data providers, it is possible to aggregate SSI estimation from multiple non-remote sensing data sources to improve the reliability of resultant SSI and solar energy estimation. Therefore, although we assume statistical parameters such as clear-sky irradiation are pre-computed as fixed coefficients, it does not affect the generalizability of our protocol. Note that our protocol can accept input data from multiple data sources, as long as the data fits into the zk-SNARK constraint system.

\section{Sonic zk-SNARK}
\label{appendix:snark_background}
In this section, we describe the basic Sonic zk-SNARK. For complete protocol descriptions, we refer the readers to  \cite{maller2019sonic}. The Sonic protocol aims to prove knowledge of three vectors $\mathbf{a, b, c} \in {\mathbb F}_p^n$, such that they satisfy the multiplication constraint of form \begin{equation}\label{eq:mul_cons}
    \mathbf a \circ \mathbf b = \mathbf c
\end{equation}
and $Q$ linear constraints of form \begin{equation}\label{eq:lin_cons}
    \mathbf a \cdot \mathbf u_q + \mathbf b \cdot \mathbf v_q + \mathbf c \cdot \mathbf w_q = k_q
\end{equation}
where $\mathbf{u_q, v_q, w_q} \in {\mathbb F}_p^n$ are coefficient vectors for the q-th linear constraint, with instance value $k_q \in {\mathbb F}_p$. All constraints are combined into one equation:
\begin{equation}
C(Y)=\mathbf a \cdot \mathbf u(Y) + \mathbf b \cdot \mathbf v(Y) + \mathbf c \cdot \mathbf w(Y) + \sum^n_{i=1}a_i b_i(Y^i + Y^{-i}) - k(Y)
\end{equation}
where $u(Y), v(Y), w(Y), k(Y)$ are polynomials derived from coefficient vectors $\mathbf{u_q, v_q, w_q}, k_q$ with indeterminate $Y\in {\mathbb F}_p$. The above polynomial is the constant term of a Laurent polynomial $t(X,Y)$ in form of:
\begin{equation}
t(X,Y) = r(X, 1)\big(r(X,Y) + s(X,Y)\big) - k(Y)
\end{equation}
where $s(X,Y)$ is a polynomial derived from constraints polynomials $u(Y), v(Y), w(Y)$. $r(X,Y)$ is a polynomial with indeterminate $X,Y$ with input vectors as coefficients: \begin{equation}
r(X,Y)=\sum^n_{i=1}a_iX^iY^i + b_iX^{-i}Y^{-i} + c_iX^{-i-n}Y^{-i-n}
\end{equation}
Given the choice of $(\mathbf a, \mathbf b, \mathbf c, k(Y))$, if the constraints are satisfied, for arbitrary $Y$ the coefficient of $X^0$ in $t$ is zero  \cite{bootle2016efficient}. 
To prove the above statement, Sonic requires a bounded extractable polynomial commitment scheme (such as the modified KZG commitment proposed in  \cite{maller2019sonic}) and a Signature of Correct Computation. After a trusted set up, the prover sends commitment of polynomial $r(X,1)$ and $t(X,y)$ and later open the commitment of $r(z,1), r(zy,1), t(z,y)$ at random challenge $z, y$ chosen by the verifier. The prover also computes and sends $s(z,y)$ with a signature of correct computation. The verifier checks and accepts the proof if and only if the following equation holds:\begin{equation}
    t(z,y) = r(z,1)\big(r(zy,1) + s(z,y)\big) - k(y)
\end{equation} 
The structured reference string is specially designed such that the constant terms in polynomials are ignored. We refer to the trusted setup of zk-SNARK scheme as choosing the bilinear group ${\sf bp}=(p,\mathbb G_1,\mathbb G_2,\mathbb G_T,e,g,h)$, and choosing random $x,\alpha$ with large enough $d$ to generate the {\sf srs}: 
\begin{equation}
    {\sf srs}_{\rm zkp}=\Big((g^{x^i}, h^{x^i}, h^{\alpha x^i})^{d}_{i=-d},(g^{\alpha x^i})^{d}_{i=-d,i\neq 0}, e(g,h^{\alpha})\Big)
\end{equation}

\section{Polynomial Commitment}
\label{appendix:commit}
A cryptographic polynomial commitment allows a prover and a verifier to agree on a polynomial while keeping the polynomial hidden and later evaluating the polynomial on specific values. A polynomial commitment scheme consists of three functions ({\sf Commit}, {\sf Open}, {\sf Verify}) defined over a bilinear group ${\sf bp}$ and a structured reference string ${\sf srs}$, where:
\begin{itemize}
    \item $F \xleftarrow{} {\sf Commit}({\sf bp}, {\sf srs}, f(X))$ takes a polynomial with indeterminate $X$ and outputs the commitment $F$. 
    \item $(f(x), \pi) \xleftarrow{} {\sf Open}({\sf bp}, {\sf srs}, F, x, f(X))$ takes a point $x$ in the field and evaluates the polynomial at the point. Additionally, it also returns a proof of evaluation $\pi$.
    \item $\{0,1\} \xleftarrow{} {\sf Verify}({\sf bp}, {\sf srs}, F, x, f(x), \pi)$ takes a pair $(x, f(x))$ and a proof $\pi$ and verifies the opening of $F$ at $x$ is $f(x)$. 
\end{itemize}
A useful polynomial commitment scheme is KZG polynomial commitment scheme  \cite{kate2010constant}. We refer to the original paper for the implementation of the commitment protocol. The trusted setup of a KZG polynomial commitment scheme involves choosing a bilinear group ${\sf bp}$ and ${\sf srs} =\Big((g^{\alpha^i})^{n}_{i=1},h^{\alpha}\Big)$. In Sonic protocol, as the polynomial commitment scheme commits to a Laurent Polynomial with degrees between $-d$ and $max$  \cite{maller2019sonic}, the {\sf srs} has a degree from $-d$ to $max$. Note that $g^{\alpha}$ is removed from ${\sf srs}$, such that it does not allow a polynomial with non-zero constant term to be committed.

\section{Setup of Sonic zk-SNARK}
\label{appendix:snark_detail}
In this section, we introduce the process of converting an index-based insurance claim into the constraint system of Sonic zk-SNARK, then set up the zk-SNARK protocol. Given the area covered by $N$ pixels, and the insurance period with $T$ sample points, the SSI calculation method introduced in Section \ref{SSI_detail} is expressed as the following Hadamard products and inner products:
\begin{align}
    (\mathbf D_{1_t}&=\mathbf f_t\circ \mathbf L_t)^T_{t=1}\\
    (\mathbf D_{2_t}&=\mathbf D_{1_t}\circ\mathbf \sigma_0)^T_{t=1}\\
    (\mathbf K_t&=(\mathbf 1-\mathbf D_{2_t}+\mathbf \sigma_1) \circ\mathbf 1)^T_{t=1} \\
    (\mathbf D_{3_t}&=\mathbf K_t\circ\mathbf G_{{\rm cs}_t})^T_{t=1}\\
    \mathbf D_4&=\sum^{T}_{t=1}\mathbf D_{3_t} \circ \mathbf 1 \label{eq:D3} \\
    \mathbf D_5&=\mathbf D_4\circ\mathbf S_{G_T}^{{-1}^T}\\
    \mathbf D_6&=\mathbf D_5\circ\mathbf G_{\rm prd}\\
    G&=\mathbf D_6 \cdot \mathbf 1 \label{eq:D6} \\
    (G_e \cdot \epsilon - G)& = \mathbf{2} \cdot \mathbf B \label{eq:G}\\
    (\mathbf{1-B}) \cdot \mathbf B & =0 \label{eq:zero}
\end{align}
where $\mathbf{2}$ is a vector $(1, 2, 2^1, 2^2, ..., 2^{M-1})$, and $\mathbf B \in \{0,1\}^M$ is the binary decomposition of $(G - G_e \cdot \sigma)$ with $M$ digits, and Eqns~(\ref{eq:G}), (\ref{eq:zero}) are range proof based on bit decomposition.

An honest prover should be able to provide valid assignment of $[\mathbf f_t,\mathbf L_t,\mathbf D_{1_t},\mathbf D_{2_t},\mathbf K_t,\mathbf D_{3_t},\mathbf D_4,\mathbf D_5,\mathbf D_6,\mathbf B,G]$ satisfying all the computations. Any adversary with invalid assignments should only be accepted by the verifier with negligible probability.

The equations above are converted into the Sonic-style constraint system mentioned above. Let $L=N(6T+3)+M+2$, first we construct input vectors $\mathbf{a,b,c}$. Let $\mathbf{a,b,c}\in\mathbb F^{L}$, each input vector can be divided into three parts (upper, middle, lower). The upper parts consist of input data only. This part will be used for data source verification:
\begin{equation}
    \mathbf a_{\rm upper}=\begin{pmatrix}
    (\mathbf f_{i})_{i=1}^T\\
    (\mathbf L_{1})_{i=1}^T\\
    \end{pmatrix}, \ 
    \mathbf c_{\rm upper}=\mathbf b_{\rm upper}=\mathbf 0^{(2TN)}
\end{equation}
where $\mathbf i^{(k)}$ represents a vector of integer $i$ of length $k$. 

The middle part contains Hadamard products in the algorithm. The multiplication constraint ensures the Hadamard products are computed correctly:
\begin{align}
\mathbf a_{\rm middle}=\ & \begin{pmatrix}
(\mathbf f_{i})_{i=1}^T\\
(\mathbf D_{1_i})_{i=1}^T\\
(\mathbf{1-D_{2_i}+\sigma_1})_{i=1}^T\\
(\mathbf K_{i})_{i=1}^T\\
\mathbf D_{4}\\
\mathbf D_{5}\\
\mathbf B\\
\end{pmatrix}
\end{align}
\begin{align}
\mathbf b_{\rm middle}=\ &\begin{pmatrix}
(\mathbf L_{i})_{i=1}^T\\
(\mathbf \sigma_{0})_{i=1}^T\\
\mathbf 1^{(TN)}\\
(\mathbf G_{{\rm cs}_t})_{t=1}^T\\
\mathbf S_{G_T}^{-1} \\
\mathbf G_{\rm prd} \\
\mathbf{1-B}\\
\end{pmatrix}, \ \ 
\mathbf c_{\rm middle}= \begin{pmatrix}
(\mathbf D_{1_i})_{i=1}^T\\
(\mathbf D_{2_i})_{i=1}^T\\
(\mathbf K_{i})_{i=1}^T\\
(\mathbf D_{3_i})_{i=1}^T\\
\mathbf D_{5}\\
\mathbf D_{6}\\
\mathbf 0^{(M)}\\
\end{pmatrix}
\end{align}

The lower parts contain other vectors or scalar values involved in computations that cannot be checked by multiplication constraints, including additions and inner products, which are formulated in linear constraints. 

\begin{align}
\mathbf a_{\rm lower}=\ &\begin{pmatrix}
G \\
G_e\cdot\epsilon-G\\
\end{pmatrix}, \  \ 
\mathbf c_{\rm lower}=\mathbf b_{\rm lower}=\mathbf 0^{(2)}
\end{align}

The full $\mathbf{a,b,c}$ are concatenations of the upper, middle and lower parts, which satisfy the multiplication constraint in Eqn.~(\ref{eq:mul_cons}). The Linear constraints check (1) addition and inner product operations within the algorithm and (2) different entries with the same values are indeed equal (copy constraints). There are in total 7 linear constraints listed in Table \ref{tab:linear_const}. $K_1$ checks all copy constraints. $\mathbf r_i\in \mathbb F^N$ are vectors of random challenges for $i\in\{1,2,...,5T+2\}$; $K_2$ checks the summation in Eqn.~(\ref{eq:D3}); $K_3$ check constants in input vectors are correct; $K_4$ checks inner product in Eqn.~(\ref{eq:D6}); $K_5$ checks left-hand side of Eqn.~(\ref{eq:G}) is correct; $K_6$ checks the right-hand side inner product in Eqn.~(\ref{eq:G}); $K_7$ checks consistency of $B$ by proving $(\mathbf{1-B}) \cdot \mathbf B=\mathbf 1$.

\begin{table*}
    \centering
    \caption{Linear constraints representing the insurance claim.}
    \begin{tabular}{|c||c|c|c|c|}
    \hline
        $q$&$k_q$&$\mathbf u_q$&$\mathbf v_q$&$\mathbf w_q$ \\
        \hline
        1 
        & $(\mathbf 1+\mathbf \sigma_1)\cdot\sum^{4T}_{3T+1} \mathbf r_{i}$ 
        & $\begin{pmatrix}
            (\mathbf -r_{i})_{i=1}^{i=2T}\\
            (\mathbf r_{i})_{i=1}^T\\
            (\mathbf -r_{i})_{i=2T+1}^{i=3T}\\
            (\mathbf r_{i})_{i=3T+1}^{i=4T}\\
            (\mathbf -r_{i})_{i=4T+1}^{i=5T}\\
            \mathbf 0^{(N)} \\
            \mathbf -r_{8T+1} \\
            \mathbf 0^{(M+2)} \\
          \end{pmatrix} $
        & $\begin{pmatrix}
            \mathbf 0^{(2NT)} \\
            (\mathbf r_{i})_{i=T+1}^{i=2T}\\
            \mathbf 0^{(3NT+2N+M+2)} \\
          \end{pmatrix}$
        & $\begin{pmatrix}
            \mathbf 0^{(2NT)} \\
            (\mathbf r_{i})_{i=2T+1}^{i=5T}\\
            \mathbf 0^{(NT)} \\
            \mathbf r_{8T+1} \\
            \mathbf 0^{(N+M+2)} \\
          \end{pmatrix} $\\
        2 
        & $0$ 
        & $\begin{pmatrix}
            \mathbf 0^{(6NT)} \\
            \mathbf -1^{(N)}\\
            \mathbf 0^{(N+M+2)} \\
          \end{pmatrix} $
        & $\begin{pmatrix}
            \mathbf 0^{(L)} \\
          \end{pmatrix}$
        & $\begin{pmatrix}
            \mathbf 0^{(5NT)} \\
            \mathbf 1^{(NT)}\\
            \mathbf 0^{(2N+M+2)} \\
          \end{pmatrix} $\\
        3 
        & $\begin{array}{l}
         \sum\limits^{6T}_{t=5T+1}\mathbf\sigma_0\cdot\mathbf r_{t}+ \sum\limits^{7T}_{t=6T+1}\mathbf r_{t}+ \sum\limits^{8T}_{t=7T+1}\mathbf G_{{\rm cs}_t}\cdot\mathbf r_{t}+ \\ 
                 \mathbf S_{G_T}^{-1}\cdot\mathbf r_{8T+2}+ 
        \mathbf G_{\rm prd}\cdot\mathbf r_{8T+3} \end{array}$
        & $\begin{pmatrix}
            \mathbf 0^{(L)} \\
          \end{pmatrix} $
        & $\begin{pmatrix}
            \mathbf 0^{(3NT)} \\
            (\mathbf r_{i})_{i=5T+1}^{i=8T}\\
            \mathbf r_{8T+2} \\
            \mathbf r_{8T+3} \\
            \mathbf 0^{(M+2)} \\
          \end{pmatrix}$
        & $\begin{pmatrix}
            \mathbf 0^{(6NT+2N)} \\
            \mathbf r_{8T+4} \\
            \mathbf 0^{(2)} \\
          \end{pmatrix} $\\
        4 
        & $0$
        & $\begin{pmatrix}
            \mathbf 0^{(6NT+2N+M)} \\
            -1 \\
            0 \\
          \end{pmatrix} $
        & $\begin{pmatrix}
            \mathbf 0^{(L)} \\
          \end{pmatrix} $
        & $\begin{pmatrix}
            \mathbf 0^{(6NT+N)} \\
            \mathbf 1^{(N)}
            \mathbf 0^{(M+2)}
          \end{pmatrix}$ \\
        5 
        & $G_e\cdot\epsilon$
        & $\begin{pmatrix}
            \mathbf 0^{(L-2)} \\
            1 \\
            1 \\
          \end{pmatrix} $
        & $\begin{pmatrix}
            \mathbf 0^{(L)} \\
          \end{pmatrix} $
        & $\begin{pmatrix}
            \mathbf 0^{(L)} \\
          \end{pmatrix}$ \\
        6 
        & $0$
        & $\begin{pmatrix}
            \mathbf 0^{(6NT+2N)} \\
            \mathbf 2\\
            0 \\
            -1 \\
          \end{pmatrix} $
        & $\begin{pmatrix}
            \mathbf 0^{(L)} \\
          \end{pmatrix} $
        & $\begin{pmatrix}
            \mathbf 0^{(L)} \\
          \end{pmatrix}$ \\
        7 
        & $M$
        & $\begin{pmatrix}
            \mathbf 0^{(6NT+2N)} \\
            \mathbf 1^{(M)}\\
            \mathbf 0^{(2)} \\
          \end{pmatrix} $
        & $\begin{pmatrix}
            \mathbf 0^{(6NT+2N)} \\
            \mathbf 1^{(M)}\\
            \mathbf 0^{(2)} \\
          \end{pmatrix} $
        & $\begin{pmatrix}
            \mathbf 0^{(L)} \\
          \end{pmatrix}$ \\
    \hline
    \end{tabular}
        \label{tab:linear_const}
\end{table*}

With the constraint system established, following the Sonic protocol we can concatenate all constraints into one polynomial. First, we sum the multiplication constraints separated by an indeterminate $Y\in\mathbb F_p$:
\begin{equation}\label{eq:mul_concat}
\sum^L_{i=1}(a_ib_i-c_i)Y^i=0
\end{equation}
and we sum the linear constraints ($Q=7$):
\begin{equation}\label{eq:lin_concat}
\sum^Q_{q=1}(\mathbf a\cdot\mathbf u_q+\mathbf b\cdot\mathbf v_q+\mathbf c\cdot\mathbf w_q-k_q)Y^{q+L}=0
\end{equation}
Then we let
\begin{align}
u_i(Y)=\ & \sum^Q_{q=1}Y^{q+L}u_{q,i}, \ \
v_i(Y)= \sum^Q_{q=1}Y^{q+L}v_{q,i}
\\
w_i(Y)=\ & \sum^Q_{q=1}Y^{q+L}w_{q,i}, \ \ 
k(Y)= \sum^Q_{q=1}Y^{q+L}k_{q}
\end{align}
Then we can combine Eqns~(\ref{eq:mul_concat}) and (\ref{eq:lin_concat}) into one polynomial of indeterminate $Y$:
\begin{align}
\mathbf a\cdot \mathbf u(Y)+\mathbf b\cdot \mathbf v(Y)+\mathbf c\cdot \mathbf w(Y)-k(Y) \notag \\ + \sum^L_{i=1} c_i \cdot (-Y^i-Y^{-i}) +\sum^L_{i=1}a_ib_i(Y^i+Y^{-i}) & \ =0
\end{align}

Based on  \cite{bootle2016efficient}, the left-hand side of the above equation can be embedded into the constant term of $t(X,Y)$ such that:
\begin{align}
t(X,Y)=\ & r(X,1)(r(X,Y)+s(X,Y))-k(Y)\\
r(X,Y)=\ & \sum^L_{i=1}a_iX^iY^i+\sum^L_{i=1}b_iX^{-i}Y^{-i}+\sum^L_{i=1}c_iX^{-i-L}Y^{-i-L}
\\
s(X,Y)=\ & \sum^L_{i=1}u_i(Y)X^{-i}+\sum^L_{i=1}v_i(Y)X^{i}+\sum^L_{i=1}w_i(Y)X^{i+L}
\end{align}

The rest of the protocol proves that the constant term of $t(X,Y)$ is zero in zero knowledge. Since ${\sf srs}$ does not allow a polynomial with non-zero constant term to be committed, all proved polynomial commitments have a zero constant term.  The interactive version of the protocol is presented below. The protocol can be converted into a non-interactive version using the Fiat-Shamir heuristic. We ignored the Signature of Correct computation and let the verifier compute $s(z,y)$ by itself. The protocol is shown in Table~\ref{tab:zksnarkProtocol}.

\section{Protocol Details}
\label{Appendix:protocol_details}
In this section, we present the full description of our PSEI insurance claim protocol. The protocol is divided into three stages: Initialization, Deployment, and Payment Settlement. 
\begin{enumerate}

\item
\textbf{Initialization}: In this stage, the insurer sets up the digital insurance contract and public parameters of the claim are chosen and shared between the insuree, contract, and RSP. We assume RSP can use an asymmetric key signature scheme $\mathcal Sig=({\sf Sign}, {\sf Ver}, {\sf Sig}_{p}, {\sf Sig}_{s})$ and a cryptographic hash function $\mathcal H$ for data validation. The public key ${\sf Sig}_{p}$ and verification function ${\sf Sig}$ are known by the insuree and smart contract. We also assume an honest RSP such that it will always provide correct data as requested by the client. However, the client may cheat by requesting data from an incorrect location or time.

\begin{enumerate}
    \item The insurer and insuree make an agreement on the insurance details, including insured area, time period, RSP, amount insured, premium, expiring time, etc. 
    \item Upon choice of insured area and RSP, the RSP setup a signature scheme and send the signature public key $\mathcal{S}_{\rm pub}$ to the contract. Our protocol adapts to common digital signature schemes such as ECDSA \cite{johnson2001elliptic}. The RSP then set up a polynomial commitment scheme by choosing a bilinear group map ${\sf bp}_{\rm RSP}$ and a structured reference string ${\sf srs}_{\rm RSP}$. Furthermore, the RSP and the contract agree on a hash function $\mathcal{H}$. These public parameters are stored on-chain. 
    \item The contract sets up the zk-SNARK scheme by independently choosing another bilinear group map ${\sf bp}_{\rm zkp}$ and structured reference string ${\sf srs}_{\rm zkp}$. Then using pre-defined SSI model parameters it can convert the SSI model into multiplication and linear constraints, and hence it deducts polynomials $s(X,Y)$ and $k(Y)$ following the construction process described in the previous section. 
\end{enumerate}

\item
\textbf{Deployment}: Following the zk-SNARK protocol introduced in the last section, the policyholder can claim the payment by proving that the claim conditions are satisfied. Notice that in Sonic protocol, $r(X,Y)$ is the only polynomial using secret input $\mathbf{a, b, c}$ as polynomial coefficients. As shown in appendix \ref{appendix:snark_detail}, vector $\mathbf{a,b}$ consists of the satellite data and other local variables. We separate $r(X,Y)$ into $r_{\rm raw}$ and $r_{\rm local}$, where $r_{\rm raw}$ contains parts of $r(X,Y)$ with $a_{\rm upper}$ as coefficients, and $r_{\rm local}$ contains the rest of $r(X,Y)$. To prove the insuree's computation using authenticated satellite imagery data provided by RSP, the polynomial $r_{\rm raw}$ is committed by RSP with signature. The claiming process is shown below:

\begin{enumerate}
    \item \textbf{Data Request}: The insuree (prover) first requests satellite data from RSP. On requesting remote sensing data at time $t$ on the insured location, the RSP creates an array $\mathbf L_t \in \mathbb {Z}^{N}$ representing radiance observed at each pixel over the location, and the pixel-wise array of calibration factor $\mathbf f_t \in \mathbb {Z}^{N}$. Here $N$ is the number of pixels covering the insured areas. The RSP then generate the Laurent polynomial commitment of the $r_{\rm raw}$:
    \begin{equation}
        r_{\rm raw}(X,1)=\sum^N_{i=1}L_{t_i}X^i+\sum^{2N}_{i=N+1}f_{t_i}X^i
    \end{equation}
    
    The commitment together with other associative information $I_{\rm aso}$ (such as geometric coordinates and timestamps) are hashed and signed by the digital signature: 
    \begin{equation}
        h = \mathcal H({\sf Commit}(r_{\rm raw})|\mathcal H(I_{\rm aso}))
    \end{equation}
    \begin{equation}
        S={\sf Sign}(h, {\sf Sig}_{s})
    \end{equation}
    Finally, the RSP responds to the data request with raw data $L_t,f_t$ and the signature $S$. Notice that because the insuree also has the raw data, it can reconstruct the $r_{\rm raw}$ polynomial and its commitment without the help of RSP. However, the signature $S$ ensures that the insuree cannot cheat by creating random $r_{\rm raw}$ using arbitrary input data. The insuree and the smart contract can validate the data by calculating $h$ and check that ${\sf Verify}(h, S, {\sf Sig}_{p})={\sf accept}$. 

    \item \textbf{Proof Generation}: The insuree then proves to the verifier that the obtained data satisfies the claiming condition by generating a proof. It is worth noticing that in the original Sonic protocol, $r(X,Y)$ is the only polynomial using secret input $\mathbf {a, b, c}$ as coefficients. As shown in Appendix \ref{appendix:snark_detail}, the secret input contains raw data. By separating raw data from other entries polynomial $r$ is divided into $r_{\rm raw}$ and $r_{\rm local}$. To validate that the raw data used to construct the proof is indeed the one provided by the RSP, the commitment and opening of $r_{\rm raw}$ are provided by the RSP and verifiable through the signature. The insuree only needs to compute commitment and openings of $r_{\rm local}$. Consequently, the proof $\pi$ is
    \begin{align*}
        \pi=\!\Big(& {\! \sf Commit}(r_{\rm local}(X,1)), {\sf Commit}(t(X,y)), r_{\rm local}(z,1),  r_{\rm local}(zy,1), \\ & \ \ {\sf Commit}(r_{\rm raw}), r_{\rm raw}(z,1), r_{\rm raw}(zy,1), t(z,y) \Big)
    \end{align*}
    where \begin{equation}
        r_{\rm local}(X,1) = r(X,1)-r_{\rm raw}(X,1)
    \end{equation}
    Applying Fiat-Shamir heuristic, the random challenges $z, y$ are generated by the client and the verifier independently, and hence we can remove steps 2 and 4 in the protocol shown in Table~\ref{tab:zksnarkProtocol}. 

    \item \textbf{Claim Verification}: Finally, the verifier checks whether the claim is valid following the protocol in Table~\ref{tab:zksnarkProtocol} step 6, with slight modifications: It first checks the validity of raw data by computing the hash message from RSP $h$ using client-provided $r_{\rm raw}$ and verifying the signature using ${\sf Sig}_{\rm pub}$. Then the verifier checks the zero-knowledge proof by verifying all opening of commitments, and the equation becomes
    \begin{align}
        t(z,y) \overset{?}{=} \ & (r_{\rm local}(z, 1)+r_{\rm raw}(z, 1)) \cdot \notag \\ & \Big(r_{\rm local}(zy, 1)+r_{\rm raw}(zy, 1) + s(z,y)\Big) - k(y)
    \end{align}
    The verifier accepts the claim if the proof is valid and proceeds to the Payment Settlement stage. Otherwise, the claim is rejected and the verifier waits for another claim. 
\end{enumerate}

\item
\textbf{Insurance Payout}: Based on the results of claim verification, the insurance payout is settled to the addresses stored in the initialization stage. If the claim is valid, the sum insured would be automatically transferred from the contract account to the insuree's account. The contract is terminated if the claim happens or the expiring date is passed. If the claim fails, the premium will be transferred to the insurer's account before the contract terminates. 

\end{enumerate}

\begin{table}[ht]
\caption{The zk-SNARK protocol based on Sonic. The interactive steps (2) and (4) can be removed by applying Fiat-Shamir heuristic.}
\begin{tabular}{|l|}
\cline{1-1}
    \parbox{\linewidth}{%
\textbf{Public Input}: info = ${\sf bp}, {\sf srs}, s(X,Y), k(Y)$,\\[-10pt]

\quad\qquad\qquad\quad commitment scheme = ({\sf Commit}, {\sf Open}, {\sf Verify})\\[-10pt]

\textbf{Prover’s Input}: $\mathbf{a, b, c}$ \\
\textbf{PSEI Insurance Claim Protocol}:
\begin{enumerate}
\item 
Prover $\xrightarrow{}$ Verifier:
\[
R={\sf Commit}({\sf bp},{\sf srs},L,r(X,1)+{\sf mix})
\]
where \[
{\sf mix}=\sum^4_{i=1}c_{i}X^{-2L-i}Y^{-2L-i}, \quad
{c_i}\xleftarrow{\$}\mathbb F_p
\]
\item
Verifier $\xrightarrow{}$ Prover: $y\xleftarrow{\$}\mathbb F_p$

\item
Prover $\xrightarrow{}$ Verifier: 
\[
T={\sf Commit}({\sf bp},{\sf srs},L,t(X,y))
\] 

\item
Verifier $\xrightarrow{}$ Prover:  $z\xleftarrow{\$}\mathbb F_p$

\item
Prover $\xrightarrow{}$ Verifier: 
\begin{align*}
a&={\sf Open}(R,z,r(X,1)) \\
b&={\sf Open}(R,yz,r(X,1)) \\ 
t&={\sf Open}(T,z,t(X,y))    
\end{align*}

\item
Verifier checks:
\[
t\overset{?}{=}a(b+s(z,y))-k(y) \wedge
\]
\[
{\sf Verify}(a,z,R) \wedge
{\sf Verify}(b,yz,R) \wedge
{\sf Verify}(t,z,T)
\]

\end{enumerate}
   }\\%
\cline{1-1}
\end{tabular}
\label{tab:zksnarkProtocol}
\end{table}

\end{document}